\documentclass[journal]{IEEEtran}
\IEEEoverridecommandlockouts
\usepackage{graphicx} 
\usepackage{amsmath, amsfonts, amssymb}
\usepackage{xcolor}
\usepackage{multirow}
\usepackage{tabularx}
\usepackage{adjustbox}
\usepackage{longtable}
\usepackage{booktabs}
\usepackage{caption}

\usepackage{algorithm}
\usepackage{algpseudocode}
\usepackage{pbalance}
\usepackage{soul}

\usepackage[flushleft]{threeparttable}

\algnewcommand\algorithmicforeach{\textbf{for each}}
\algdef{S}[FOR]{ForEach}[1]{\algorithmicforeach\ #1\ \algorithmicdo}

\newcommand{\LL}[1]{\text{log}_2 \left(#1 \right)}
\newcommand{\Ll}[1]{\text{log}_2 #1}

\title{H-FA: A Hybrid Floating-Point and Logarithmic Approach to Hardware Accelerated FlashAttention}
\author{Kosmas Alexandridis and Giorgos Dimitrakopoulos
\thanks{%
This work was supported by an Industry Cooperation funded by Infineon Technologies Austria AG in the course of IPCEI Microelectronics.

Kosmas Alexandridis and Giorgos Dimitrakopoulos are with the Department
of Electrical and Computer Engineering, Democritus University of Thrace, Xanthi, Greece.
E-mail: \{koalexan, dimitrak\}@ee.duth.gr}}

\begin{document}

\maketitle

\begin{abstract}
Transformers have significantly advanced AI and machine learning through their powerful attention mechanism. However, computing attention on long sequences can become a computational bottleneck.
FlashAttention mitigates this by fusing the softmax and matrix operations into a tiled computation pattern that decouples performance from sequence length. Though designed for GPUs, its simplicity also makes it well suited for direct hardware acceleration. To improve hardware implementation, we compute FlashAttention using a mixture of floating-point and fixed-point logarithm domain representations.
Floating-point is used to compute attention scores from query and key matrices, while logarithmic  computation simplifies the fused computation of softmax normalization and the multiplication with the value matrix. 
This transformation, called H-FA, replaces vector-wide floating-point multiplication and division operations by additions and subtractions implemented efficiently with fixed-point arithmetic in the logarithm domain. Exponential function evaluations are effectively omitted and fused with the rest operations, and the final result is directly returned to floating-point arithmetic without any additional hardware overhead.
Hardware implementation results at 28nm demonstrate that H-FA achieves a 26.5\% reduction in area and a 23.4\% reduction in power, on average, compared to FlashAttention parallel hardware architectures built solely with floating-point datapaths, without hindering performance.
\end{abstract}


\section{Introduction}
Modern ML and AI systems rely on deep learning models with billions of parameters, reaching human-level performance in tasks like image recognition~\cite{vit} and natural language processing~\cite{deepseek}. Central to this progress is the attention mechanism~\cite{base_attn}, which enables models to selectively focus on relevant parts of the input, leading to powerful architectures such as Transformers. These models are built from layers combining multi-head self-attention and feed-forward networks, supported by positional encodings to retain word order. Multi-head attention allows the model to capture diverse relationships across the input by processing information in parallel subspaces. With residual connections and layer normalization, Transformers effectively model long-range dependencies and now form the backbone of Large Language Models (LLMs).

Attention compares user queries with a set of key vectors, producing
an attention score matrix that determines how much emphasis
to place on each value vector when generating a response.
The query, key, and value vectors all originate from the same
set of embeddings but are transformed using different learned
weight matrices. Specifically, attention computes $\text{softmax}(QK^T)V$, that includes a matrix multiplication $QK^T$ across query $Q$ and key $K$ matrices, softmax normalization to obtain attention weights, and a final matrix multiplication with value matrix $V$.

As Transformers process increasingly longer sequences, the attention mechanism emerges as a major computational bottleneck~\cite{longformer}. Handling extended contexts in standard attention raises quadratically the number of
operations and memory use. 
To address this, sparse, linear, and low-rank attention~\cite{sparse_attn, lin_attn, low_rank_attn} have been proposed, which aim to reduce complexity by either simplifying the attention matrix or narrowing focus to the most informative parts of the input. These approaches seek a trade-off between maintaining model accuracy and improving efficiency.

Hardware accelerators for attention focus on managing data locality by keeping queries, keys, and values, in on-chip SRAM~\cite{a3,keller,lu,mecla}. When on-chip storage is insufficient to hold intermediate results across full sequences, data must be written back to external memory, creating performance bottlenecks. These external memory accesses can be reduced by compressing the Transformer's weights~\cite{kim}, however,
to completely overcome this sequence-length dependency, designs such as~\cite{lazy_softmax,cosa, moon,chen} enhance efficiency by optimizing core attention operations including matrix multiplications and softmax evaluations. Beyond optimizing data movement and computation order, other approaches leverage token similarity~\cite{elsa,tsacc,fact} to avoid redundant operations, reducing both latency and energy use. Other strategies, like in-memory computation~\cite{xformer}, aim to streamline attention processing by performing computation directly within memory arrays.

FlashAttention~\cite{fa, fa2, nsquared}, initially designed for GPUs, has become a highly effective method for speeding up attention computations. It achieves this by combining softmax evaluation with matrix operations in a tiled fashion, enabling efficient parallelism and minimizing memory access overhead. This integrated approach not only shortens execution time but also scales well with longer sequences. As a result, FlashAttention overcomes key performance bottlenecks without sacrificing accuracy of attention outputs.

In this work, we first argue that the simplicity of FlashAttention is not only beneficial for GPUs but also enables the design of efficient parallel hardware accelerators for attention. Second, we move part of FlashAttention’s computations into the logarithmic domain, significantly reducing hardware cost by (a) converting vector-wise multiplications and divisions into additions and subtractions, and (b) avoiding explicit exponential evaluations by leveraging the fused exponential and multiplication operations inherent to the FlashAttention algorithm. 
For the latter case, a similar approach, but with a narrower scope, has been proposed in~\cite{isvlsi-25}. The key advantages of the proposed hardware architecture, named \emph{H-FA}, are summarized as follows:

\begin{itemize}
\item
H-FA implements FlashAttention-2 using a combination of floating-point and fixed-point logarithmic computations.
Floating-point arithmetic is used to compute attention scores from the query and key matrices, while the fused softmax normalization and multiplication with the value matrix are performed in the logarithmic domain. This approach replaces the vector-wide multiplications and softmax divisions in FlashAttention-2 with simpler additions and subtractions in the log domain. Explicit evaluations of the exponential function are eliminated by fusing them into the surrounding operations.
\item
The reduced dynamic range of logarithmic representation allows the use of fixed-point arithmetic, significantly reducing hardware cost compared to floating-point implementations. Most importantly, the applied logarithmic quantization maintains accuracy for representative LLM applications, as demonstrated by inference on Phi-3.5-mini-instruct~\cite{phi}, Llama-3.2~\cite{llama3} and Qwen2~\cite{qwen2} models, using various benchmarks for LLM evaluation.
\item 
To evaluate the effectiveness of H-FA in executing FlashAttention, we implemented a parallel hardware dataflow accelerator for both the optimized FlashAttention-2 kernel and the proposed H-FA architecture. Both designs were synthesized down to physical layout using 28nm ASIC technology. Experimental results show that H-FA, which employs a hybrid of floating-point and logarithmic computation in carefully selected parts of the FlashAttention algorithm, achieves substantial area and power savings, ranging from 22.5\% to 27\% across various head dimension sizes, while maintaining equal execution time.
\end{itemize}
The rest of the paper is organized as follows: Section~\ref{s:attn} revisits Attention and FlashAttention algorithm. Section~\ref{s:flash-hw} discusses the parallel hardware implementation of FlashAttention. Section~\ref{s:proposed} introduces the proposed hybrid float-log domain computation of FlashAttention and its efficient hardware implementation. Experimental results are presented in Section~\ref{s:eval} and conclusions are drawn in the last Section.
\section{Computation of Attention and FlashAttention}
\label{s:attn}

In attention, the model computes a similarity score between a query vector and a set of key vectors. These scores are used to generate a weighted sum of value vectors, determining how much focus to place on each value. The query, key, and value vectors are all derived from the same input but are transformed using different learned weight matrices.

\subsection{Attention Kernel}
For a query $\vec{q}$ and 
a set of key and value vectors 
$K=\vec{k}_1,\ldots, \vec{k}_N$ and 
$V=\vec{v}_1, \ldots, \vec{v}_N$ attention is defined as 
\begin{equation}
s_i = \text{dot}(\vec{q}, \vec{k}_i)\qquad f_i = \frac{e^{s_i}}{\sum_j e^{s_j}}\qquad \text{Attn}(\vec{q}, \text{K}, \text{V}) = \sum_i f_i\, \vec{v}_i
\nonumber
\end{equation}

The attention score $s_i$ quantifies how similar a given query is to the $i$th key vector, calculated using a dot product. To determine which tokens are most relevant to a specific query, all associated scores undergo a softmax transformation. This process involves exponentiating each score and then normalizing by dividing with the total sum of these exponentials. The resulting attention weights are then used to compute a weighted sum over the value vectors, where each value's influence on the final output is based on its corresponding attention weight. 

\begin{algorithm}[t]
\caption{Attention with lazy softmax division}
\label{alg:attn-lazy}
\begin{algorithmic}[1]
\ForEach {query $\vec{q}$}
\For{$i = 1:N$} 
\State $s_i \gets \text{dot}(\vec{q}, \vec{k}_i)$
\State $m_i \gets \max(m_{i-1}, s_i)$
\EndFor
\State $\ell_0 \gets 0$
\For{$i = 1:N$} 
\State $\vec{o}_i \gets \vec{o}_{i-1} + e^{s_i - m_N}\cdot \vec{v}_i$
\State $\ell_i \gets \ell_{i-1} +e^{s_i - m_N}$
\EndFor
\State $\text{attn}(\vec{q}, K, V) \gets \vec{o}_N/ \ell_N$
\EndFor
\end{algorithmic}
\end{algorithm}

In practice, attention scores are scaled by $\sqrt{d}$, where $d$ is the head dimension size, prior to applying softmax normalization~\cite{base_attn}. Additionally, some attention scores are masked (i.e., set to a large negative value) to prevent attending to certain positions, such as padding tokens or future tokens in causal attention~\cite{base_attn}. To simplify the algorithmic derivation, we omit these features from the following descriptions.

Exponentiating large scores can lead to numerical instability or infinite values. To mitigate this, the \emph{safe softmax} technique is used: it subtracts the highest score from all the individual scores before applying the exponential, maintaining the function’s behavior while preventing overflow, i.e., 
\begin{equation}
f_i = \frac{e^{s_i-\max}}{\sum_j e^{s_j-\max}} \quad \text{(safe softmax)}
\nonumber
\end{equation}

Rather than computing the exponential sum and performing normalization before combining with the value vectors, lazy softmax architectures~\cite{lazy_softmax, elsa} take a different approach. They simultaneously accumulate both the weighted sum of the value vectors and the sum of the exponentials, deferring the division step until the very end of the process.
\begin{equation}
s_i\!=\! \text{dot}(\vec{q}, \vec{k}_i)\quad f_i\!=\! e^{s_i-\max}\quad 
\text{Attn}(\vec{q}, \text{K}, \text{V})\!=\! \frac{\sum_i f_i \vec{v}_i}{\sum_j e^{s_j-\max}}\nonumber
\end{equation}
Alg. 1 illustrates how attention is computed using the lazy softmax division approach~\cite{lazy_softmax, elsa}. Initially, it calculates the dot product between the query vector and each key vector. During this step, the maximum score is also identified, which is essential for applying a numerically stable softmax. Next, the output vector $\vec{o}_i$ is built incrementally by summing the product of each value vector and its corresponding exponentiated score adjusted by subtracting the maximum score. Simultaneously, the cumulative sum of these exponentials, $\ell_i$, is tracked. To obtain the final attention vector for the query $\vec q$, the accumulated output is divided by the total sum of exponentials $\ell_N$.

\subsection{FlashAttention}
FlashAttention~\cite{fa}, building on the principles of online softmax computation~\cite{online-softmax}, reformulates attention as a streaming process. The enhanced version, FlashAttention-2~\cite{fa2} that also delays softmax division until the final stage~\cite{lazy_softmax, nsquared}, is shown in Alg.~\ref{alg:flash-attn2}.
Unlike the standard lazy softmax method shown in Alg.~\ref{alg:attn-lazy}, FlashAttention-2 performs all intermediate computations including score calculation, exponentiation, and accumulation, within a single inner loop. This avoids the separate step of finding the global maximum score, making it particularly well-suited for handling long input sequences, where efficiency becomes increasingly critical.
 
\begin{algorithm}[t]
\caption{FlashAttention-2 with delayed softmax division}\label{alg:flash-attn2}
\begin{algorithmic}[1]
\ForEach {query $\vec{q}$}
\For{$i = 1:N$} 
\State $s_i \gets \text{dot}(\vec{q}, \vec{k}_i)$
\State $m_i \gets \max(m_{i-1}, s_i)$
\State $\ell_i \gets \ell_{i-1}e^{m_{i-1}-m_i}+e^{s_i-m_i}$
\State $\vec{o}_i \gets \vec{o}_{i-1} e^{m_{i-1}-m_i}+\vec{v}_i e^{s_i-m_i}$
\EndFor
\State $\text{attn}(\vec{q}, K, V) \gets \vec{o}_N/\ell_N$
\EndFor
\end{algorithmic}
\end{algorithm}

In each iteration of Alg.~\ref{alg:flash-attn2}, the similarity score $s_i$ is obtained by computing the dot product between the query vector and a key vector. The current maximum score $m_i$ is then identified. The running sum of exponentials, $\ell_i$, is updated by adding $e^{s_i - m_i}$. 
When a new maximum $m_i$ exceeds the previous one $m_{i-1}$, the previous accumulated value $\ell_{i-1}$ is rescaled by a factor of $e^{m_{i-1} - m_i}$, maintaining consistency across iterations. The output vector $\vec{o}_i$ is updated in a similar fashion: the current value vector $\vec{v}_i$, weighted by its adjusted exponential score, is added to the rescaled previous output $\vec{o}_{i-1} e^{m_{i-1} - m_i}$. Once all iterations are complete, the final output vector $\vec{o}_N$ is normalized by dividing it by the total accumulated exponent sum $\ell_N$, yielding the final attention result for the given query.

\section{FlashAttention-based Hardware Accelerators}
\label{s:flash-hw}
In this work, we argue that the structured and streamlined design of the FlashAttention-2 algorithm makes it an excellent candidate for implementation on dedicated hardware accelerators. To achieve substantial reductions in execution time, such an accelerator must support parallel computation of attention.

\subsection{Parallel FlashAttention units}
As shown in Alg.~\ref{alg:flash-attn2}, the FlashAttention-2 kernel includes two nested loops, both of which offer opportunities for unrolling to boost parallelism and computational efficiency. Although unrolling the inner loop is feasible, it retains serial dependencies across the internal state variables, $m_i$, $\ell_i$, and $\vec{o}_i$, which limits scalability. 

\begin{figure}[t]
\centering
\includegraphics[width=0.98\columnwidth]{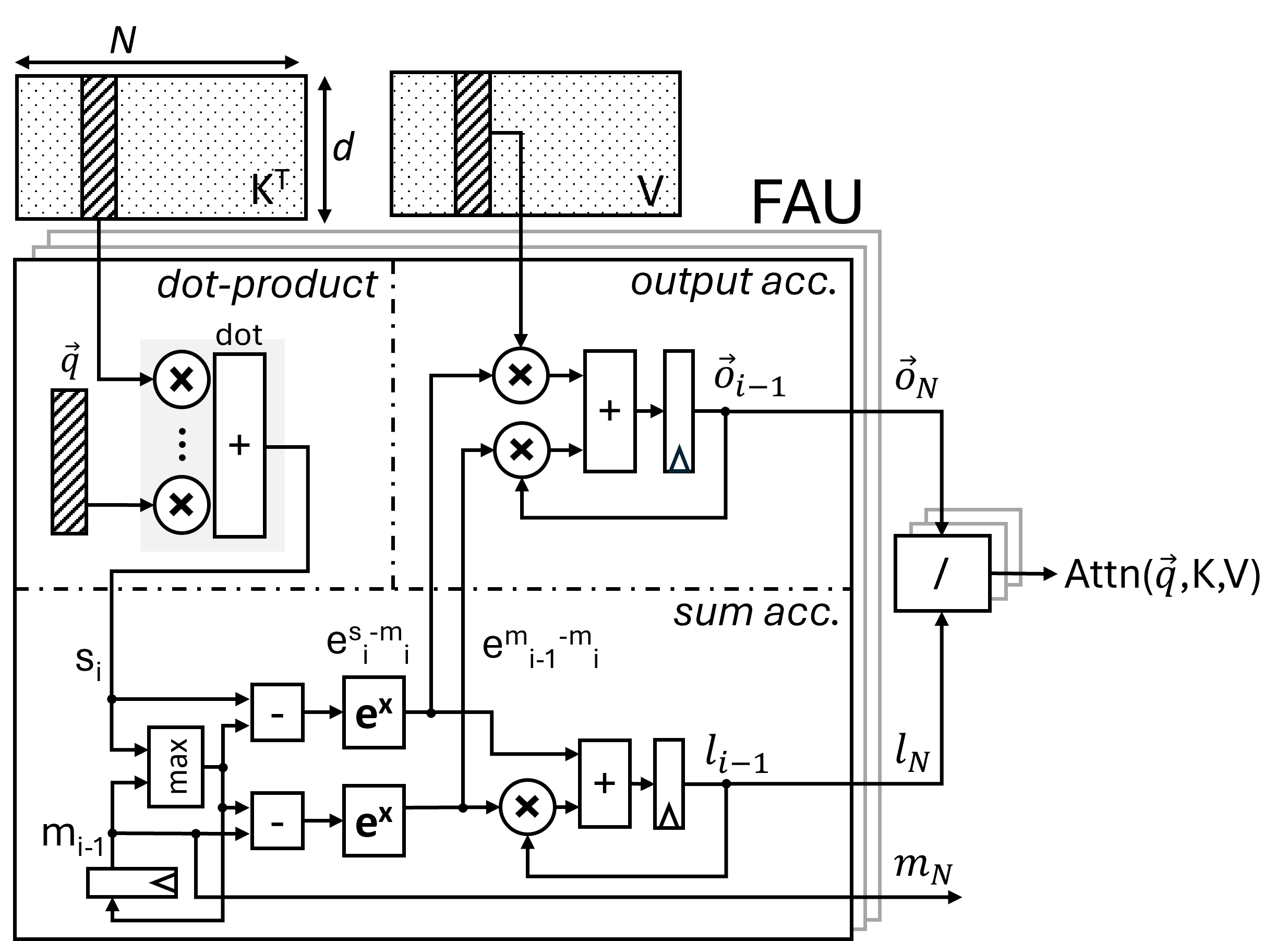}
\caption{The organization of the FlashAttention Unit (FAU) serving one query vector. Multiple FAUs can serve multiple query vectors in parallel.}
\label{f:flashattn2-hw}
\end{figure}

A more effective strategy is to unroll the outer loop, enabling the kernel to process multiple query vectors concurrently \emph{reusing the same blocks of key and value vectors}. This approach maintains a separate internal state for each query, removing the serialization bottleneck. The corresponding parallel hardware design is illustrated in Fig.~\ref{f:flashattn2-hw}.
The core hardware block servicing each query vector is the FlashAttention Unit (FAU) that consists of three distinct blocks: The \emph{dot-product} unit, the \emph{sum accumulator} and the \emph{output accumulator}.
The `dot-product' unit of each FAU computes the similarity of the query vector with each key vector . 
This results in an updated maximum score and a revised sum of exponentials for each query computed in the `sum acc.' unit.
In parallel, value vectors enter the `output acc.' unit of the FAU, updating the output vector.
Once all keys and values have been processed, a final division step completes the attention computation for each query vector. The procedure concludes after all queries in the batch have been evaluated.

FlashAttention-2 avoids the requirement for a full softmax hardware module by decoupling the exponentiation and final division steps, allowing each to be computed independently. Various hardware-friendly techniques have been explored to implement these non-linear operations efficiently. These include piecewise linear approximations applied after range reduction~\cite{koca}, methods based on logarithmic quantization~\cite{sole}, and alternative approximations~\cite{exp2two} that express exponentials as powers of two, making them compatible with simple shift-and-add logic that leads to fast and low-cost hardware.

\begin{figure}
\centering
\includegraphics[width=0.9\columnwidth]{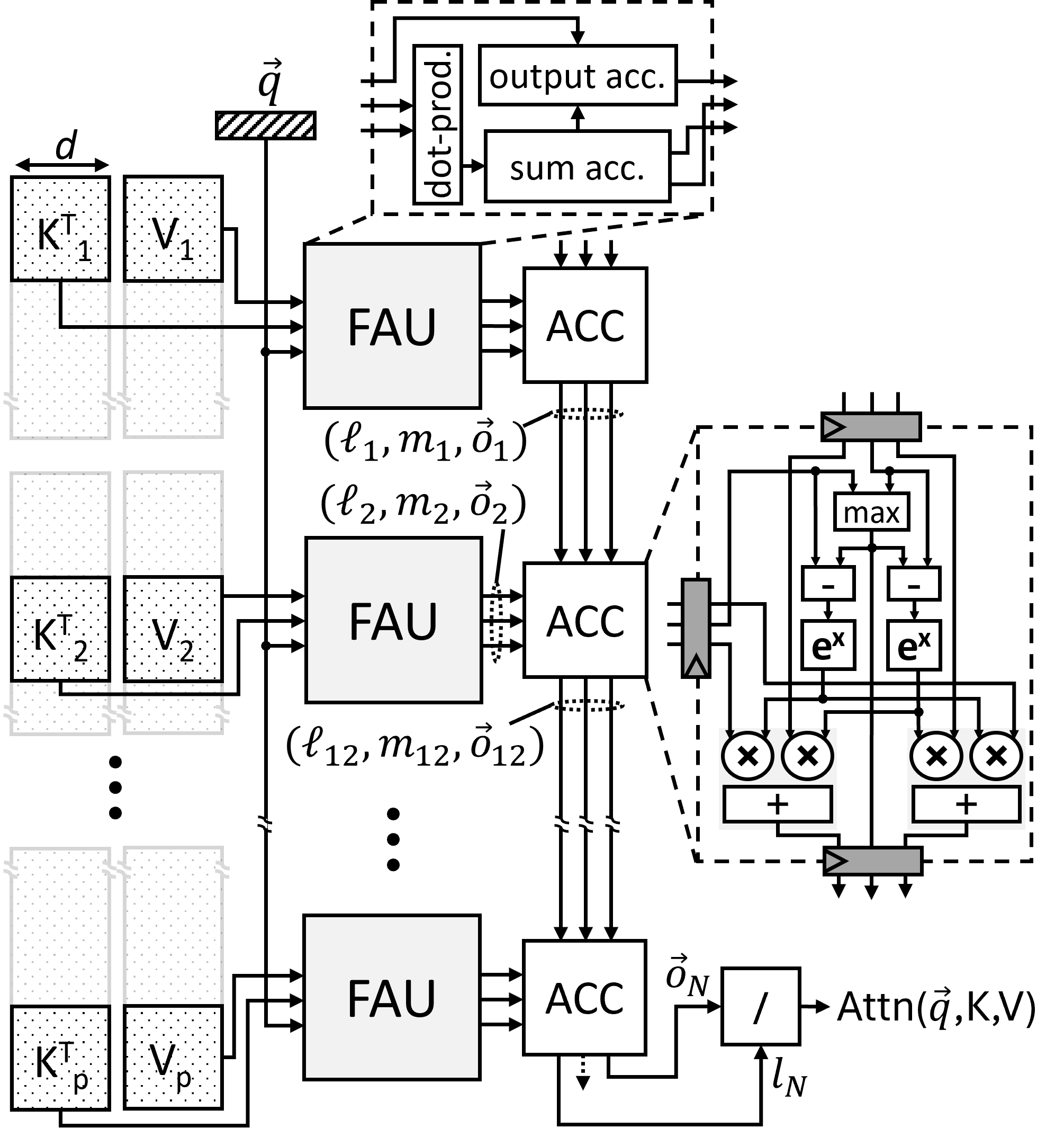}
\caption{Block-level computation of attention for a single query vector. The set of key and value vectors are split into multiple blocks which are computed in parallel by the FlashAttention units (FAUs). The partial attention outputs are accumulated in the vertical direction using cascaded ACC units that implement Eq.~\eqref{e:merge}. The final normalized attention value is computed using division.}
\label{f:flashattn2-hw-2d}
\end{figure}

\subsection{FlashAttention Units operating in parallel to multiple KV sub-blocks}
The parallelism available for computing attention can be further exploited for a single query vector by adopting 2D attention techniques~\cite{fa2, block, attention2d}. In this approach, the attention output for a given query vector can be computed concurrently across different blocks of the key-value matrices, as long as the intermediate results from each block are appropriately scaled and combined in an online manner.

Specifically, consider the inner loop of Alg.~\ref{alg:flash-attn2}, excluding the final normalization step in line 8. Suppose this loop is executed in parallel across two blocks of the $K$ and $V$ matrices, denoted as $(K_A, V_A)$ and $(K_B, V_B)$. These blocks represent the two halves of the full $K$ and $V$ matrices, respectively. For the same query vector $\vec{q}$, each block independently computes a partial attention output.
To reconstruct the final output vector $\vec{o}_N$, the partial results $(\vec{o}_A, \ell_A)$ and $(\vec{o}_B, \ell_B)$, produced by the two halves, must be merged using the following equations, as described in \cite{attention2d, lean-attn}:
\begin{align}
m_N &= \max(m_A, m_B) \nonumber \\ 
\vec{o}_N &= \vec{o}_A\,e^{m_A - m_N} + \vec{o}_B\,e^{m_B - m_N} \label{e:merge} \\ 
\ell_N &= \ell_A\,e^{m_A - m_N} + \ell_B\,e^{m_B - m_N} \nonumber
\end{align}
Here, $m_A$ and $m_B$ represent the maximum attention scores computed independently within each sub-block.

Fig.~\ref{f:flashattn2-hw-2d} shows the parallel, block-level computation of the attention vector for a single query block. The query vector $\vec{q}$ is broadcast to multiple blocks, with each block handling a subset of the key-value matrix' rows. Each FAU, for each sub-block, independently computes a partial result triplet $(\ell_i, m_i, \vec{o}_i)$, which are then combined into the final output using cascaded ACC blocks that perform the merging described in Eq.~\eqref{e:merge}.
The FAU in each block is identical to the one in Fig.~\ref{f:flashattn2-hw}, except that the final normalization step (division) is deferred until all $KV$ blocks have been processed.

The operation of the parallel attention computation block is divided into two phases.
In the first phase, all block-FAU units work in parallel to accumulate their respective $(\ell_i, m_i, \vec{o}_i)$ triplets, each accessing all key–value vectors within a KV block. The time required for this phase is proportional to the depth of each KV block that equals to $N/p$, where $N$ refers to sequence length and $p$ to the number of parallel FAU units.
In the second phase, the ACC units complete the accumulation of intermediate attention vectors through the vertical pipeline. To ensure synchronization between the two phases and smooth data transfer along the accumulation pipeline, the block-FAUs and ACC units communicate via a ready/valid pipelined flow-control protocol. Each ACC unit begins its computation once the outputs from both the block FAU and the preceding ACC are valid.

Parallel attention computation significantly reduces the time required to compute attention for a single query block, assuming the KV sub-blocks are either preloaded into local buffers or that memory bandwidth is sufficient to stream them efficiently during computation.

Effectively, the designs shown in Figs.~\ref{f:flashattn2-hw} and~\ref{f:flashattn2-hw-2d} allow the computation of attention on multiple queries in parallel, where the attention for each query is computed independently for multiple $KV$ matrix sub-blocks, and the same $KV$ sub-blocks are reused for all queries.  
\section{Logarithmic-Domain Optimization of FlashAttention Kernel}
\label{s:proposed}
The primary objective of this work is to apply logarithmic computations only in places that would have the most impact in simplifying the overall hardware complexity.
In FlashAttention, we perform computations in logarithmic domain
to eliminate the expensive exponentiation, multiplication, and division operations in lines 5 and 6 and 8 of Alg.~\ref{alg:flash-attn2}.
Specifically, we will show that all intermediate accumulations for $\vec{o}_i$ and $\ell_i$ can be performed \emph{entirely in the log domain}, using only a single conversion back to the linear domain at the final stage to recover the true attention output from its logarithmic representation.

Dot product between query and key vector as well as the computation of maximum attention scores and their differences in lines 3, 4 and 5 of Alg.~\ref{alg:flash-attn2} are computed in floating point without any further change.

This selective transformation of FlashAttention allows us to bypass expensive exponent–multiply operations, as logarithm arithmetic enables these operations to be fused into simpler additions and shifts. Moreover, the final division step, required to normalize the attention output for each query, becomes a  subtraction in the log domain.

\subsection{Sum of two products in Logarithm Number System}
In this section, we briefly revisit how a sum of two products
\begin{equation}
c = xz + yu
\label{e:lin-dp}
\end{equation}
is computed in the logarithmic number system (LNS)~\cite{log-cnn}. This formulation will be used in the next section to compute $\ell_i$ and $\vec{o}_i$ that appear in lines 5 and 6 of Alg.~\ref{alg:flash-attn2}.

An arbitrary floating point number $x\in \mathbb R$ can be represented in LNS with a tuple of its sign and the base-2 logarithm of its absolute value, as follows:
\begin{equation}
\text{LNS}: (s_x, X) \leftrightarrow x = (-1)^{s_x}\cdot 2^{X},\ \text{with}\ X=\log_2{|x|} 
\end{equation}
The sign bit $s_x$ is $0$ when $x$ is positive and $1$ otherwise. $X$ follows a fixed point representation with appropriately selected integer and fraction bits. 

Using the LNS representations
$(s_x, X)$, $(s_z, Z)$, $(s_y, Y)$, and $(s_u, U)$
for the variables $x$, $z$, $y$, and $u$, respectively, we can compute $c$ in~\eqref{e:lin-dp} as follows:
\begin{align}
c = xz + yu & = (-1)^{s_x+s_z}\, 2^{X+Z} + (-1)^{s_y+s_u}\,2^{Y+U} \nonumber\\
& =  (-1)^{s_{xz}}\,2^{X+Z} + (-1)^{s_{yu}}\,2^{Y+U} 
\label{e:c-1}
\end{align}
where $s_{xz} = s_{x} \oplus s_z$ and $s_{yu} = s_{y} \oplus s_u$, with $\oplus$ referring to the exclusive-OR boolean function.

The LNS representation of $c$ requires the computation of $\log_2|c|$ and separately its corresponding sign. Depending on signs $s_{xz}$ and $s_{yu}$ and the values of $X+Z$ and $Y+U$ the $\log_2|c|$ and its sign $s_c$ are equal to
\begin{align}
\log_2|c| & = \begin{cases}
\log_2|2^{X+Z}+2^{Y+U}|, \quad s_{xz} = s_{yu} \\
\log_2|2^{X+Z}-2^{Y+U}|, \quad s_{xz} \ne s_{yu} \\
\end{cases} \label{e:logc}\\
    s_c &= \begin{cases}
        s_{xz}, & X+Z > Y+U\\    
        s_{yu}, & Y+U \ge X+Z
    \end{cases} 
\end{align}
Effectively $\log_2|c| = \log_2|2^{X+Z}\pm 2^{Y+U}|$ and the actual operation, i.e., addition or subtraction is determined by the signs $s_{xz}$ and $s_{yu}$.

Defining $A = X + Z$ and $B = Y + U$ and assuming that $A > B$,  $\log_2|c|$ can be rewritten as follows:
\begin{align}
    \log_2|c| & = \log_2|2^A\pm 2^B| \nonumber\\
             &= \Ll{|2^A (1 \pm 2^{B-A})|} \nonumber  \\ 
             &= \Ll{|2^A|} + \Ll{|1 \pm 2^{B-A}|} \nonumber  \\
             &= A + \Ll{|1 \pm 2^{B-A}|} \label{e:ge-a}
\end{align}
Similarly when $A \le B$ 
\begin{equation}
    \Ll{|c|} = B + \Ll{|1 \pm 2^{A-B}|} \label{e:ge-b}
\end{equation}
Combining~\eqref{e:ge-a} and~\eqref{e:ge-b}, we can write the expression for $\Ll{|c|}$, in a more compact form as:
\begin{equation}
    \Ll{|c|} = \text{max}(A,\ B) + \LL{1 \pm 2^{-|A-B|}} \label{e:comp-lns-add}
\end{equation}
According to~\cite{lns-add}, based on the signs of $A$ and $B$, i.e., $s_A = s_{xz}$ and $s_B=s_{yu}$, we decide how to compute the LNS representation of $c$:
\begin{align}
    \Ll{|c|} &= \begin{cases}
        \text{max}(A,\ B) + \LL{1 + 2^{-|A-B|}}, s_{xz} = s_{yu} \\    
        \text{max}(A,\ B) + \LL{1 - 2^{-|A-B|}}, s_{xz} \ne s_{yu}    
    \end{cases} \nonumber \\ 
    s_c &= \begin{cases}
        s_{xz}, & A > B\\    
        s_{yu}, & B \ge A
    \end{cases} 
    \label{e:lns-add}
\end{align}

\subsection{Attention output calculation in logarithmic domain}

The initial step in simplifying the FlashAttention-2 kernel involves combining two related computations: the progressive accumulation of the sum-of-exponents (originally in line 5 of Alg.~\ref{alg:flash-attn2}) and the incremental scaled update of the output vector (line 6 of the same algorithm). Since both processes share similar computational patterns, they can be unified into
a single vectorized update:
\begin{equation}
\begin{bmatrix}
\ell_i \\           
\vec{o}_i
\end{bmatrix}=
\begin{bmatrix}
\ell_{i-1}\cdot e^{m_{i-1}-m_i}+1\cdot e^{s_i-m_i} \\           
\vec{o}_{i-1}\cdot e^{m_{i-1}-m_i}+\vec{v}_i\cdot e^{s_i-m_i}
\end{bmatrix}
\label{e:merged}
\end{equation}
By appending one additional element to both the output vector $\vec{o}_i$ and the value vector $\vec{v}_i$, i.e., defining 
$O_i = [\ell_i\quad \vec{o}_i]$ and $V_i = [1\quad \vec{v}_i]$, 
the combined incremental update for the sum-of-exponents and the output, as shown in Eq.~\eqref{e:merged}, can be reformulated as:
\begin{equation}
O_i = O_{i-1}\, e^{m_{i-1}-m_i} + V_i\, e^{s_i-m_i}
\label{e:merged-output}
\end{equation}
The final values of the output vector $\vec{o}_N$ and the accumulated sum of exponents $\ell_N$, which are needed for the normalization step in line 8 of Alg.~\ref{alg:flash-attn2}, can be extracted directly from $O_N$. 

Since we target custom hardware implementations, it is more preferable to work with powers of two rather than natural exponents. Therefore, knowing that $e^x=2^{x\log_2 e}$, we can write~\eqref{e:merged-output} as follows:
\begin{equation}
O_i = O_{i-1}\, 2^{(m_{i-1} - m_i)\log_2e} + V_i\, 2^{(s_i-m_i)\log_2e} 
\label{e:basic}
\end{equation}

The merged output accumulation~\eqref{e:basic} is a sum of two products that can be computed entirely in the log domain after appropriate data reformatting and using the formulation of the previous Section summarized in Eq.~\eqref{e:lns-add}.

While for $O_{i-1}$ and $V_i$ we need to derive their equivalent representation in LNS including their sign bits, the terms $2^{(m_{i-1} - m_i)\log_2e}$ and $2^{(s_i - m_i)\log_2e}$ are always positive and can be considered to be already in logarithmic form as long as the exponents $(m_{i-1} - m_i)\log_2e$ and $(s_i - m_i)\log_2e$ are quantized to the same fixed-point format used by the LNS representation of the rest two. This observation effectively minimizes how many variables actually need to pay the cost of moving from linear to the logarithm domain. 
Overall, the computation of~\eqref{e:basic} can be performed in the log domain according to~\eqref{e:lns-add} as follows:
\begin{subequations}
\label{e:lns-attn}    
\begin{align}
    \label{e:lns-attn-a}
    \Ll{|O_i|} &= \text{max}(A, B) + \LL{1 \pm 2^{-|A-B|}}\\
    \label{e:lns-attn-b}
    A &= \Ll{|O_{i-1}|} + \text{quant}[(m_{i-1}-m_i)\log_2e] \\
    \label{e:lns-attn-c}
    B &= \Ll{|V_i|} + \text{quant}[(s_i-m_i)\log_2e] \\
    \label{e:lns-attn-d}
    s_{O_i} &= \begin{cases}
        s_{O_{i-1}} &,\ \text{if}\ A > B \\
        s_{V_i} &,\ \text{if}\ B \ge A
    \end{cases}
\end{align}
\end{subequations}

The form of Eq.~\eqref{e:lns-attn}, which expresses the logarithm of the updated output $\log_2 |O_i|$ entirely in terms of the previous value $\log_2 |O_{i-1}|$, highlights a key target of this work: the accumulation of both the true output vector $\vec{o}$ and the associated sum of exponents $\ell$ inside $O$ can be carried out \emph{entirely} within the logarithmic domain, without requiring intermediate conversions back to the linear domain. Actually, the only variable that needs to be transformed from linear to the log domain before being used is the value vector $V_i$.

This simplification is made possible by our decision to apply the logarithmic domain only to the second part of the FlashAttention computation. The computation of the attention scores $s_i$, as well as the current and previous maximum attention scores $m_i$ and $m_{i-1}$, respectively, remains in the linear domain.
The quantization of attention score differences $m_{i-1}-m_i$ and $s_i-m_i$ as required in~\eqref{e:lns-attn-b} and~\eqref{e:lns-attn-c} occurs in a small dynamic range.
Since $m_i$ is defined as the maximum of $m_{i-1}$ and $s_i$, both differences under quantization are guaranteed to be negative. In the linear domain, these differences are used to compute exponential terms $e^{m_{i-1} - m_i}$ and $e^{s_i - m_i}$ in~\eqref{e:merged-output}, which quickly approach zero when their exponents fall below approximately $-15$. Based on this observation, we constrain both differences to the range 
$[-15,0]$, enabling their quantization using a uniform 16-bit fixed-point format with 9 integer bits and 7 fractional bits, as required by the LNS format chosen for the rest variables.

As long as output $O_i$ is kept entirely in its logarithmic form, i.e., $\log_2 |O_i|$, the division step in line 8 of Alg.~\ref{alg:flash-attn2} can be carried out directly in the logarithmic domain as well.
Given that $O_N = [\ell_N \quad \vec{o}_N]$, the attention output can be expressed in LNS as:
\begin{align}
\label{e:fin-res-lns}
\log_2|\text{Attn}(\vec{q}, K, V)| &= \log_2 (|\vec{o}_N/\ell_N|) \nonumber \\ &= \log_2 |\vec{o}_N| - \log_2 |\ell_N| \\
s_{\text{Attn}(\vec{q}, K, V)} &= s_{\vec{o}_N} \oplus s_{\ell _N}\nonumber
\end{align} 
As long as both the final output vector $\vec{o}_N$ and the sum of exponents $\ell_N$ are maintained exclusively in the log domain, this subtraction yields the logarithm of the final attention result without requiring any intermediate conversion of datatypes. The only necessary transformation occurs at the very end, when the final attention result is converted back to the linear domain.

\subsection{Attention accumulation from multiple KV sub-blocks in the log domain}
In case that multiple FAUs are instantiated and operate in parallel to multiple KV sub-blocks, as shown in Fig.~\ref{f:flashattn2-hw-2d}, we need to accumulate the partial attention results to a final attention output. To avoid redundant conversions between floating point and logarithmic fixed point, we transfer the computation of attention accumulation in the log domain as well. This is performed using exactly the same methodology described in the previous paragraphs.
For instance, the partial attention merging $o_N = o_Ae^{m_A-m_N} + o_Be^{m_B-m_N}$ given in~\eqref{e:merge} is computed in the log domain as follows:
\begin{subequations}
\label{e:lns-acc}
\begin{align}
    \label{e:lns-acc-a} 
    \Ll{|\vec{o}_N|} &= \text{max}(A, B) + \LL{1 \pm 2^{-|A-B|}} \\
    \label{e:lns-acc-b}
    A &= \Ll{|\vec{o}_A|} + \text{quant}[(m_A-m_N)\log_2e] \\
    \label{e:lns-acc-c}
    B &= \Ll{|\vec{o}_B|} + \text{quant}[(m_B-m_N)\log_2e] \\
    \label{e:lns-acc-d}
    s_{\vec{o}_N} &= \begin{cases}
        s_{\vec{o}_A} &,\ \text{if}\ A > B \\
        s_{\vec{o}_B} &,\ \text{if}\ B \ge A 
    \end{cases}
\end{align}
\end{subequations}
\noindent A similar derivation is followed for merging the sum-of-exponents $\ell_N$ of~\eqref{e:merge}.

\section{Logarithmic FlashAttention Unit in Hardware}
\label{s:log_attn}
In this section, we will describe how~\eqref{e:lns-attn} and effectively~\eqref{e:lns-acc-a}
are computed in hardware highlighting all the needed intermediate details. Additionally, we detail the transformation of the final logarithmic attention score, obtained through~\eqref{e:fin-res-lns}, back to the linear domain.

\subsection{Logarithmic scaled output accumulation}
First we remove in~\eqref{e:lns-attn-a} the need to compute logarithm by following Mitchell's approximation~\cite{mitchell}. Since, by definition $0 < 2^{-|A-B|} < 1$, Mitchell has shown in~\cite{mitchell} that $\log_2(1\pm 2^{-| A-B|})\approx \pm 2^{-| A-B|}$. Therefore, $\log_2 |O_i|$ in~\eqref{e:lns-attn-a} can be equivalently written as:
\begin{equation}
\log_2 |O_i| \approx \max(A, B) \pm 2^{-| A-B|}
\label{e:out_log_acc}
\end{equation}

\noindent To compute~\eqref{e:out_log_acc} in hardware, we follow three distinct steps:

\subsubsection{Compute $A$} Following~\eqref{e:lns-attn-b}, this step involves 
adding $\log_2|O_{i-1}|$ to the quantized $(m_{i-1} - m_{i})\log_2 e$. The multiplication with $\log_2 e$ takes place in fixed point after quantizing the attention score difference.
Since the output vector $O$ remains entirely in the logarithmic domain no additional translation step is required beside the quantization of attention score differences. 

\subsubsection{Compute $B$} \label{ss:log2fx}
According to~\eqref{e:lns-attn-c} this step involves adding the logarithm of the value vector $V_i$ to the quantized $(s_i - m_{i})\log_2 e$.
The fixed-point logarithm of the floating-point value vector 
$V_i$ is computed using the method proposed in~\cite{blinn} and further examined in~\cite{sweden}.

Let $V_{ik}$ denote the $k$th element of the value vector $V_i$. Since the absolute value of $V_{ik}$ is a positive floating-point number, with exponent $E_V$ that includes the bias term $b$, and mantissa $M_V$, its logarithm can be expressed as follows:
\begin{align}
\log_2|V_{ik}| &= \LL{2^{E_V-b}\cdot(1+M_V)} \nonumber \\
&= E_V - b + \LL{1 + M_V} \nonumber \\
&\approx E_V - b + M_V
\label{e:fp-to-fx}
\end{align} 
The approximation $\log_2(1 + M_V) \approx M_V$ is based on Mitchell’s method~\cite{mitchell}, which holds since $M$ is a fractional value between 0 and 1.

Exponent $E_V$ is an integer and mantissa $M_V$ contains only fractional bits. Thus, there is no need to explicitly compute their sum $E_V+M_V$ that appears in~\eqref{e:fp-to-fx}. This can be performed implicitly~\cite{blinn} by considering $E_V$ and $M_V$ as parts of the fixed point number   $E_V.M_V$. 
The bits dedicated to the integer and fraction part correspond to the original bit width of $E_V$ and $M_V$. For instance, 
assuming that $V_{ik}$ follows BFloat16 representation, $E_V.M_V$ would have 8 integer and 7 fraction bits, respectively. To account for negative logarithm values, using signed (two's complement) arithmetic, that arise when $V_{ik} < 1$, we add one extra bit to the integer part resulting in total of 9 integer bits.

To compute $\log_2|V_{ik}|$ following~\eqref{e:fp-to-fx}, we need to subtract the bias $b$ from the integer part of $E_V.M_V$. This is performed after aligning with a left shift the bias term to the integer part.

\subsubsection{Computing $2^{-|A - B|}$}
Up to now, we have shown how $A$ and $B$ are computed using only fixed point arithmetic.  
Thus, $|A - B|$ is also a fixed point result formed by the sum of its integer part $p$ and its fraction part $f$, i.e., 
\begin{align}
2^{-|A - B|}=2^{-(p+f)}=2^{-p}\cdot2^{-f} \nonumber
\end{align}
Since $p$ is a positive integer its negative power-of-two function $2^{-p}$ is equivalent to a right-shift operation. Thus, 
\begin{align}
    2^{-|A - B|}=2^{-f} \gg p  
    \label{e:exp_ap}
\end{align}
To compute $2^{-f}$ efficiently in hardware, we employ a piecewise-linear (PWL) approximation that exploits the fact that $f$, being a fractional value, is always confined to the range 
$[0,1)$. This interval is divided into 8 uniform segments using a PWL fitting tool~\cite{pwlf}, which minimizes the approximation error within each segment. The resulting set of linear coefficients is stored in lookup tables (LUTs), which are indexed using the most significant bits of the fractional input $f$.

\subsection{Transforming fixed-point logarithm attention to a final floating-point attention result}
\label{s:h-fa2-hw}
Attention computation is finalized in~\eqref{e:fin-res-lns}, where the logarithm of the attention for one query vector is computed via a fixed-point subtraction of two logarithms that replaces costly floating-point division.
After that, we need to transform the logarithmic attention to linear domain in order to perform downstream floating-point operations. 

To perform the needed transformation, first, we write the fixed point result $\Ll{|\text{Attn}(\vec q, K, V)|}$ as a sum of its integer $I$ and fractional $F$ part
\begin{equation}
\label{e:part-log-attn}
     \log_2|\text{Attn}(\vec q, K, V)| = I+F
\end{equation}
Since $F$ is a small number between $0$ and $1$, we can utilize Mitchell's approximation as in Eq.~\eqref{e:fp-to-fx} but in the opposite direction to get that 
\begin{equation}
     \log_2|\text{Attn}(\vec q, K, V)| \approx I+\log_2(1+F)
\end{equation}
Exponentiating both sides to remove the logarithm from attention we get
\begin{equation}
\label{e:abs-attn}
|\text{Attn}(\vec q, K, V)| = 2^{I+\log_2(1+F)} = 2^I \cdot (1+F)
\end{equation}
The right-hand side of~\eqref{e:abs-attn} resembles the structure of a floating-point number, but without the bias in the exponent. To address this limitation, we add a bias term to $I$.
Moreover, since $I$ can be either positive or negative, we adjust the sign of the final result accordingly, based on the outcome of~\eqref{e:lns-attn-d}.

\begin{figure}[t]
\centering
\includegraphics[width=0.98\columnwidth]{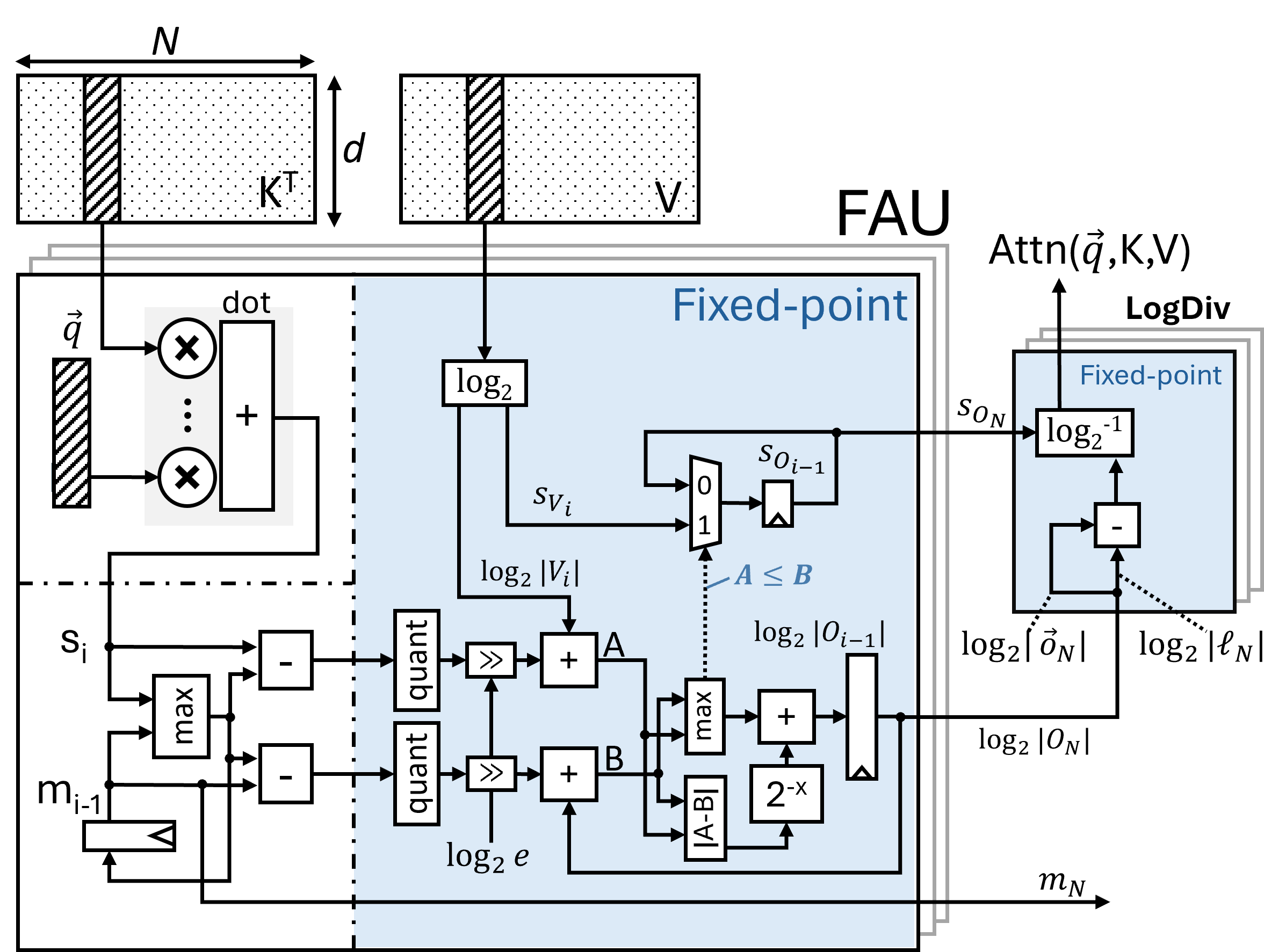}
\caption{The optimized structure of the FAU operating partly in the logarithm domain.}
\label{f:logiflash-hw}
\end{figure}

\begin{figure}[t]
\centering
\includegraphics[width=0.75\columnwidth]{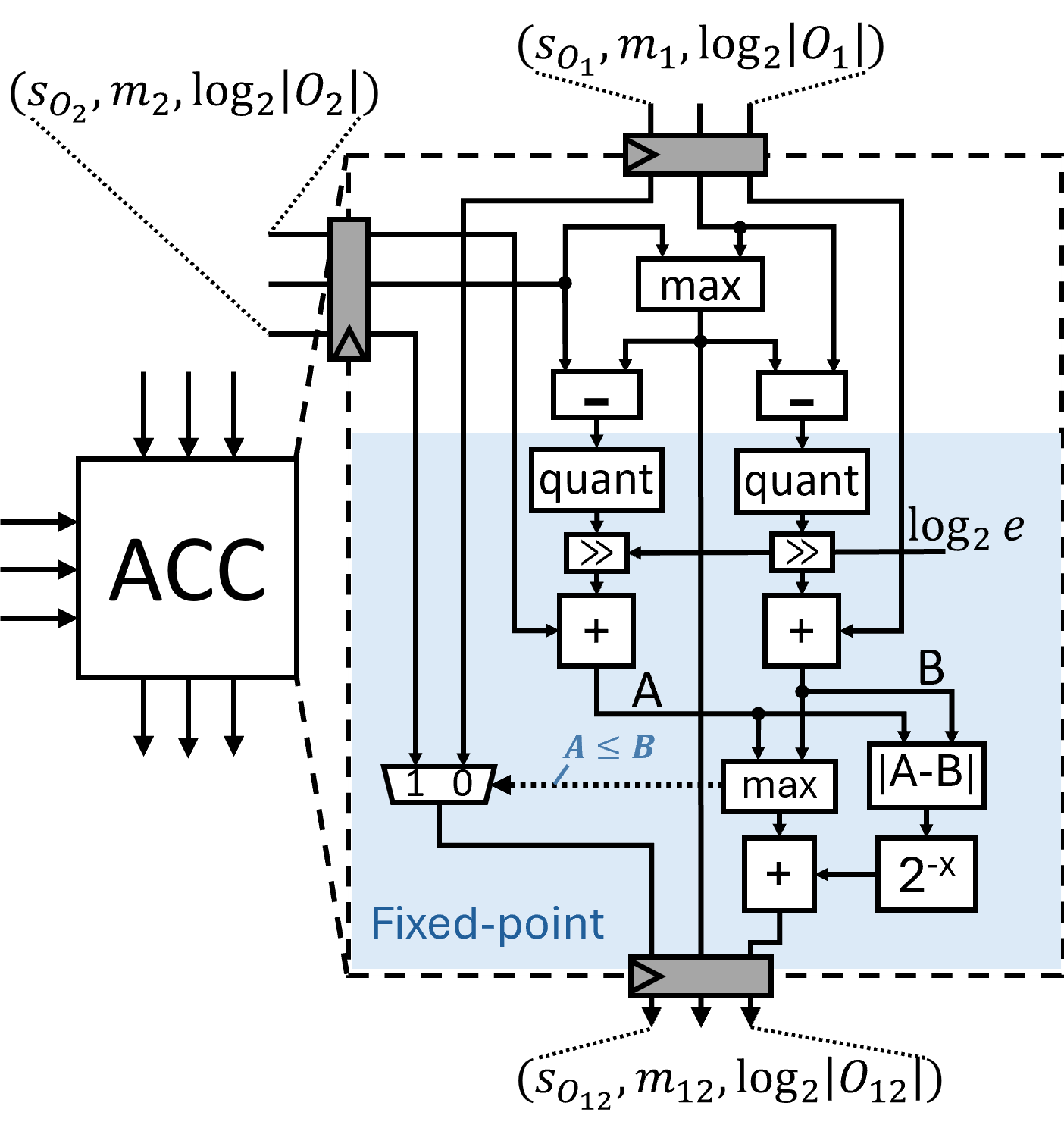}
\caption{Hardware architecture of optimized ACC blocks for accumulating partial attention results.}
\label{f:logiAcc-hw}
\end{figure}

\subsection{Organization of logarithmic FAUs and ACC blocks}
Fig.~\ref{f:logiflash-hw}, shows the optimized structure for the FAU of Fig.~\ref{f:flashattn2-hw}. 
While the dot product and the normalized softmax inputs remain in floating point, the sum accumulation of $\ell$ and output vector accumulation logic of $\vec o_N$ are fused into a single unified stage, which is performed entirely in logarithmic domain using fixed-point arithmetic.
The quantization units located at the west-side as well as the constant shifters that follow, are two for each normalized softmax input, while the rest of the fixed-point logic is replicated for each element of the value and output vector, respectively.

Conversions from and to the logarithmic domain require reinterpreting or manipulating the underlying bits. In the proposed architecture, this is needed \emph{only} for the value vectors $V_i$ and the final attention vector computed for each query. 
The final attention vector is computed inside LogDiv unit which replaces division with a fixed-point subtraction in the logarithmic domain. Since the input of LogDiv is already in the log domain only the transformation back to linear floating-point domain is needed.

Similarly, Fig.~\ref{f:logiAcc-hw} shows the hardware organization of the optimized ACC block. In this case the input and output partial result triplets contain the maximum value $m_i$, the sign of the output vector $s_i$ and the logarithm of its absolute value $\log_2|O_i|$. Note that only $m_i$ is a floating-point number while the rest are fixed-point integers. Since ACC blocks do not need to perform division in logarithmic domain they only contain quantization units to transform the max differences $m_A-m_N$ and $m_B-m_N$ to fixed--point arithmetic and do not require any conversion back to linear domain. Every operation, except the maximum value calculation and the maximum differences, is performed using fixed-point arithmetic.

\section{Evaluation}
\label{s:eval}

Experimental evaluation aims to quantify the impact of the new hybrid float-log computation of FlashAttention on real machine-learning applications and also highlight the achieved hardware savings.

\subsection{Verification of LLM accuracy}
Since part of the computation of FlashAttention is done in the logarithmic domain, it effectively introduces quantization and approximation in multiple arithmetic operations. Our goal is quantify how much these approximations affect the performance of real LLM applications that use FlashAttention kernel.
To do so, in the first set of experiments, we emulated the hybrid attention computation kernel 'H-FA' in Python and integrated it to Microsoft's Phi-3.5-mini-instruct model. Then, we executed inference tasks on the MMLU benchmark suite~\cite{mmlu} using various prompts and utilities provided by the {\tt lm-evaluation-harness} framework~\cite{lm-eval}. We simply run inference using the default weights available on the HuggingFace repository~\cite{hf} \emph{as is} and \emph{without} applying any fine-tuning or re-training.

\begin{table}[t]
\caption{Inference performance of Microsoft's Phi-3.5-mini-instruct LLM model for the MMLU benchmarks using floating-point only FlashAttention and the proposed hybrid float-log approach.}
\label{t:mmlu}
\begin{minipage}[t]{.5\columnwidth}
    \centering
    \begin{adjustbox}{width=1\columnwidth}
    \begin{tabular}[t]{|l||c|c|}
    \hline
        \multirow{2}{*}{\textbf{Benchmark}} & \multicolumn{2}{c|}{Accuracy (\%)} \\ \cline{2-3}
         ~& \textbf{H-FA} & \textbf{FA-2} \\
         \cline{1-3}
        \noalign{\vskip 2.5pt}
        \cline{1-1}
        \textbf{Other} & \multicolumn{2}{c}{} \\ \hline
        clinical knowledge & 77 & 77 \\ 
        miscellaneous & 80 & \textbf{81} \\ 
        moral disputes & 66 & \textbf{68} \\ 
        nutrition & \textbf{77} & 72 \\ 
        human aging & 66 & \textbf{67} \\ 
        human sexuality & 77 & 77 \\ 
        international law & \textbf{83} & 82 \\ 
        jurisprudence & 81 & \textbf{82} \\ 
        logical fallacies & 85 & 85 \\ 
        machine learning & 52 & \textbf{53} \\ 
        management & 81 & 81 \\ 
        marketing & 88 & 88 \\ 
        medical genetics & 75 & \textbf{76} \\ 
        moral scenarios & 60 & \textbf{63} \\ 
        philosophy & \textbf{75} & 74 \\ 
        prehistory & 76 & 76 \\ 
        public relations & \textbf{68} & 66\\ 
        security studies & 77 & 77  \\ 
        sociology & \textbf{79} & 78 \\ 
        us foreign policy & 82 & 82 \\ 
        virology & 50 & 50\\ 
        world religions & 73 & 73 \\ 
        abstract algebra & 44 & \textbf{46} \\ 
        anatomy & \textbf{64} & 60 \\ 
        astronomy & 76 & \textbf{80} \\ 
        business ethics & 74 & 74 \\ 
        computer security & \textbf{82} & 81 \\ 
        conceptual physics & 67 & \textbf{71} \\ 
        econometrics & \textbf{63} & 62 \\ 
        electrical engin. & 58 & \textbf{62} \\ 
        elementary math. & \textbf{53} & 48 \\ 
        formal logic & 57 & \textbf{58}\\ 
        global facts & 32 & \textbf{39} \\ \hline
    \end{tabular}
    \end{adjustbox}
\end{minipage}
    \begin{minipage}[t]{.49\columnwidth}
    \centering
    \begin{adjustbox}{width=1\columnwidth}
      \begin{tabular}[t]{|l||c|c|}
        \hline
        \multirow{2}{*}{\textbf{Benchmark}} & \multicolumn{2}{c|}{Accuracy (\%)} \\ \cline{2-3}
         ~& \textbf{H-FA} & \textbf{FA-2} \\ 
        \cline{1-3}
        \noalign{\vskip 2.5pt}
        \cline{1-1}
        \textbf{Professional} & \multicolumn{2}{c}{} \\ \hline
        accounting & 55 & \textbf{54} \\ 
        law & 53 & 53 \\ 
        medicine & \textbf{63} & 62 \\ 
        psychology & 72 & 72 \\
        \cline{1-3}
        \noalign{\vskip 2.5pt}
        \cline{1-1}
        \textbf{College} & \multicolumn{2}{c}{} \\ \hline  
        computer science & 54 & \textbf{57} \\ 
        medicine & 71 & \textbf{72} \\ 
        biology & 82 & \textbf{83} \\ 
        chemistry & 43 & \textbf{52} \\ 
        mathematics & 41 & 41 \\ 
        physics & 40 & \textbf{43} \\
        \cline{1-3}
        \noalign{\vskip 2.5pt}
        \cline{1-1}
        \textbf{High school} & \multicolumn{2}{c}{} \\ \hline 
        biology & 78 & \textbf{80} \\ 
        chemistry & 53 & \textbf{57} \\ 
        comp. science & \textbf{69} & 67 \\ 
        european history & 77 & \textbf{78} \\ 
        gov. and politics & 90 & 90 \\ 
        macroeconomics & \textbf{80} & 78 \\ 
        mathematics & \textbf{39} & 35 \\ 
        physics & 53 & 53 \\ 
        psychology & \textbf{92} & 90 \\ 
        statistics & \textbf{69} & 59 \\ 
        geography & \textbf{81} & 80 \\ 
        microeconomics & 85 & \textbf{86} \\ 
        us history & 88 & 88 \\ 
        world history & 80 & 80 \\
        \hline
      \end{tabular}
      \end{adjustbox}
\end{minipage}%

\end{table}

Table~\ref{t:mmlu} compares the accuracy of two Phi-3.5-mini-instruct model variants: 'H-FA', using the proposed hybrid float-log FlashAttention, and 'FA-2', using the default scaled dot-product attention (SDPA) from the Torch library, commonly used in HuggingFace LLMs.
In both implementations, all floating-point operations are performed using the BFloat16 data type. For each LLM application, the approach that achieves the best accuracy is highlighted in bold.

In 18 out the 57 LLM applications the accuracy achieved by both approaches is exactly the same.
In some cases, the 'H-FA' outperforms its full-precision counterpart. This phenomenon is a known artifact of lower-precision models~\cite{i-bert}. While it does not indicate that the model is inherently superior, it affirms the fact that the approximations used do not degrade performance. Even in cases that 'H-FA' has less accuracy the accuracy difference is small and below 5\% in the majority of cases. Only in 'global facts' and 'college chemistry' accuracy degrades more.

\begin{table}[!t]
\caption{Inference performance of LLMs utilizing floating-point only FlashAttention and the proposed hybrid float-log approach.}
    \label{t:llms_all}
\begin{threeparttable}
    \centering
    \begin{adjustbox}{width=1\columnwidth}
    \begin{tabular}{|l||l|c|c|c|c|c|}
    \cline{3-7}
        \multicolumn{2}{c|}{} & \multicolumn{5}{c|}{\textbf{Benchmark Accuracy (\%)}} \\ \cline{3-7}
        \multicolumn{2}{c|}{} & GPQA & MMLU & SWAG & GSM8K & XCOPA \\ \hline
        \multirow{2}{*}{Qwen2-0.5B} & FA-2 & 27 & 45 & 49 & 33 & 51 \\
        ~ & H-FA & 25 & 45 & 48 & 30 & 50 \\ \hline \hline
        \multirow{2}{*}{Llama3.2-1B} & FA-2 & 25 & 38 & 58 & 43 & 51 \\
        ~ & H-FA & 23 & 38 & 56 & 44 & 50 \\ \hline \hline
        \multirow{2}{*}{Phi3.5-4B$^a$} & FA-2 & 31 & 69 & 59 & 72 & 52 \\
        ~ & H-FA & 32 & 68 & 63 & 68 & 52 \\ \hline
    \end{tabular}
    \end{adjustbox}
    \begin{tablenotes}
      \small
      \item $^a$The Phi3.5-mini-instruct variant with 3.82B parameters
    \end{tablenotes}
    \end{threeparttable}
\end{table}

Inference performance was evaluated in a similar way for other contemporary benchmark LLM applications and models.
Specifically, we tested Llama3.2-1B and Qwen2-0.5B to analyze how different model sizes perform across four benchmarks: SWAG~\cite{swag}, XCOPA~\cite{xcopa}, GPQA~\cite{gpqa} and GSM8K~\cite{gsm8k}.

Table~\ref{t:llms_all} presents the accuracy achieved by each LLM on every benchmark. Since each benchmark consists of multiple sub-tasks, as shown in detail for MMLU in Table~\ref{t:mmlu}, we report the mean accuracy score for each case in Table~\ref{t:llms_all}.
Across all evaluations, the largest observed error is 4\%, occurring with the Phi3.5-4B model on GSM8K and SWAG. In all other cases, the error remains below 3\%.
These results demonstrate that the robustness of our H-FA hybrid computation method is not limited to a specific LLM or benchmark, 
but also extends to weaker models that are sensitive to numerical inaccuracies, across diverse benchmarks as reported in~\cite{weakllms}.

\subsection{Discussion of Approximation-Induced Errors}
To better understand why H-FA and its arithmetic approximations, designed to simplify the hardware implementation of FlashAttention, have minimal impact on the overall performance of the evaluated LLMs, we conducted additional experiments. H-FA introduces error at three stages of the attention computation:
(a) 
Quantization of floating-point attention score differences into fixed-point representation after applying the \textit{quant} operation in~\eqref{e:lns-attn} and~\eqref{e:lns-acc};
(b) 
Mitchell’s approximation in~\eqref{e:out_log_acc} and~\eqref{e:fp-to-fx};
(c)
The PWL approximation of the $2^{-x}$ function for fractional inputs in~\eqref{e:exp_ap}.

Our first goal was to quantify the contribution of each approximation to the total error in the output logits (i.e., the intermediate outputs of LLM attention layers). To this end, we ran multiple inferences on the same LLM and benchmark, systematically eliminating one source of error at a time. Results were averaged across all layers at each inference step to capture the total induced error, and are reported in Table~\ref{t:err_contib}.

The percentages in Table~\ref{t:err_contib} clearly show that Mitchell’s approximation is by far the dominant source of error, contributing more than 90\% of the total, while both the PWL approximation and the use of fixed-point arithmetic each account for less than 10\%.

\begin{table}[!ht]
    \caption{ Total absolute error contribution (\%) of the three error sources for three LLM/Benchmark pairs.}
    \label{t:err_contib}
    \centering
    \begin{tabular}{|c||c|c|c|}
    \cline{2-4}
        \multicolumn{1}{c}{} & \multicolumn{3}{|c|}{\textbf{Absolute Error contribution (\%)}} \\ \hline 
        \textbf{LLM/} & \multirow{2}{*}{BF16-to-FIX16} & \multirow{2}{*}{Mitchell} & \multirow{2}{*}{PWL $2^{-x}$} \\ 
        \textbf{Benchmark} &  &  & \\ \hline
        Qwen2-0.5B/ & \multirow{2}{*}{5.5} & \multirow{2}{*}{92.0} & \multirow{2}{*}{2.5} \\
        SWAG &  & &  \\ \hline
        Llama3.2-1B/ & \multirow{2}{*}{8.4} & \multirow{2}{*}{90.1} & \multirow{2}{*}{0.5} \\ 
        XCOPA & &  &  \\ \hline
        Phi3.5-4B/ & \multirow{2}{*}{7.4} & \multirow{2}{*}{92.4} & \multirow{2}{*}{0.2} \\ 
        MMLU & & & \\ \hline
    \end{tabular}
\end{table}

Having established that Mitchell’s approximation is the primary source of error in H-FA, our next step was to investigate why this approximation does not significantly degrade the overall accuracy of LLM applications.

\begin{figure}[!h]
\centering
\includegraphics[width=0.9\linewidth]{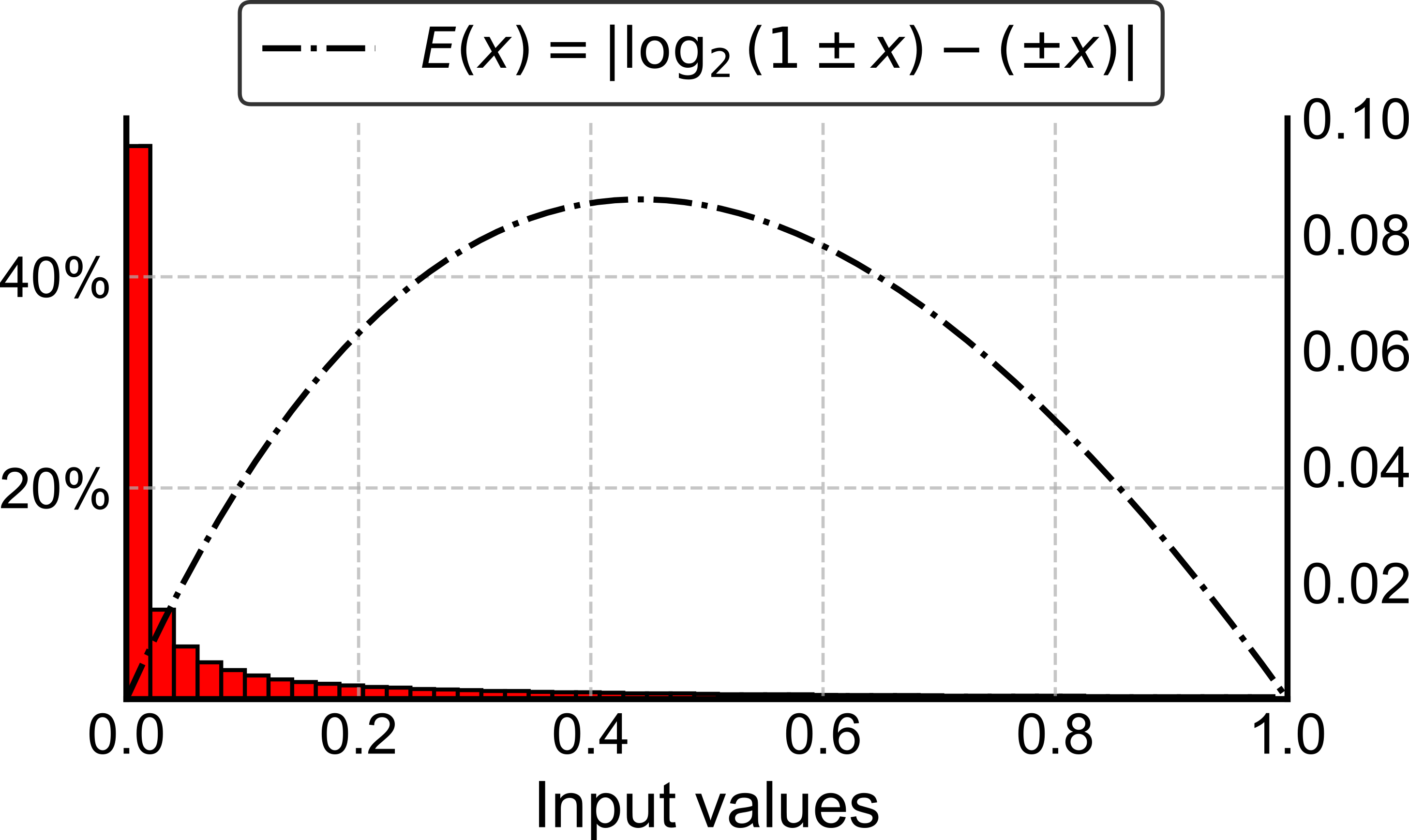}
\caption{The distribution of input values (their absolute value) undergoing Mitchell’s approximation, along with the corresponding absolute error introduced in each case. By definition, these inputs fall within the interval [0, 1].}
\label{f:val_dist}
\end{figure}

As discussed in Section~\ref{s:log_attn}, H-FA employs Mitchell’s approximation in two cases:
a) to compute the $\log_2|V|$ operation (Eq.~\eqref{e:fp-to-fx}); and
b) to simplify LNS addition (Eq.~\eqref{e:out_log_acc}).
In both cases, the approximation replaces $\log(1 \pm x)$ with $\pm x$ for $x \in [0,1]$. The error of this approximation reaches its maximum when $x$ lies near the middle of the input range. To verify how often this situation occurs in practice, we recorded all inputs where Mitchell’s approximation was applied across every LLM/benchmark pair and inference step. The resulting histogram of these recorded values are shown in Fig.~\ref{f:val_dist}, along with the absolute approximation error defined as $E(x) = |\log(1 \pm x) - (\pm x)|$. Such approximation is employed for the floating-point value vectors in~\eqref{e:fp-to-fx} whose mantissa is by definition between 0 and 1 and the simplification of logarithm of $1\pm 2^{-|A-B|}$ in~\eqref{e:out_log_acc} since $2^{-|A-B|}$ lies within the same interval.

As shown in Fig.~\ref{f:val_dist}, the histogram indicates that the vast majority of inputs subjected to Mitchell’s approximation have absolute values below 0.1.
In such cases, the absolute approximation error shown on the secondary y-axis is less than 0.02. In cases that inputs fall roughly in the center of their input range a larger error is induced that may affect the final accuracy of the LLM application. This is the reason why 'global facts' and 'college chemistry' applications for Phi-3.5-mini-instruct LLM, as discussed in the previous subsection, exhibit higher accuracy loss than other cases.

Moreover, Mitchell’s approximation has a strict upper bound: the absolute error can never exceed 0.08, and even this maximum occurs only for a small fraction of inputs. Also, this error does not accumulate across successive attention computations. 
Even in the worst-case scenario that we apply two consecutive Mitchell approximations, where both inputs correspond to the maximum error of 0.08, the resulting error remains 0.08, and not the sum of the two errors (0.16).

Taken together, these observations explain why Mitchell’s approximation, although responsible for over 90\% of the total error, can introduce at most an error of 0.08 in the final output. 

Therefore, our hybrid H-FA attention computation method does not significantly impact LLM performance, as confirmed by the reported benchmark results.

\subsection{Evaluation of hardware complexity}
To assess the area and power savings achieved by the proposed approach, we implemented two variants of a FlashAttention accelerator that computes attention using KV sub-blocks, based on the overall architecture shown in Fig.~\ref{f:flashattn2-hw-2d}. The first design, referred to as 'FA-2', represents the state-of-the-art FlashAttention accelerator that performs all computations using floating-point arithmetic. In contrast, 'H-FA' denotes the proposed design, which adopts a hybrid approach, executing part of the computation in floating point and part using fixed-point logarithmic arithmetic. In both cases, all floating-point computations refer to the BFloat16 datatype.

We implemented both designs in C++ and used Siemens EDA Catapult HLS to synthesize them into Verilog, targeting a 28-nm standard-cell technology. To ensure functional correctness, the C++ implementations were integrated into {\tt llama2.c}~\cite{llama-repo}, where they produced identical responses to the original implementation across all tested queries. Both accelerators were constrained to operate at a clock frequency of 500 MHz with identical pipelined latency. The total latency varies with the hidden dimension size, requiring 19, 20, and 21 cycles for $d = {32, 64, 128}$, respectively. For computing the dot products of a query and a key vector, multi-operand floating-point addition follows the architecture presented in~\cite{ol-align}.

Physical synthesis of Verilog RTL was carried out using Cadence's digital implementation tools, and power estimates of the datapaths under comparison were obtained using the Siemens EDA PowerPro analysis and optimization suite. The reported power reflects average consumption measured during inference on various benchmarks using the {\tt lm-evaluation-harness} framework. The area and power contribution of SRAM memory for implementing KV buffers is also included in the evaluation. For the attention accelerator we assume that it supports a maximum sequence length of $N=1024$ tokens. Thus, each key and value matrix consists of 1024 rows that are distributed to four blocks of 256 rows each. Each row contains  $d$ BFloat16 elements (following the size $d$ of the hidden dimension).
The area and power of KV SRAM buffers were estimated using Cacti~\cite{cacti} through Accelergy's hardware components Python library {\tt hwcomponents}~\cite{accelergy} at $22$nm technology node. Acquired results were up-scaled to the $28$nm node by utilizing the DeepScale tool~\cite{deepscale}.

\begin{figure}[t!]
    \centering
    \includegraphics[width=0.5\columnwidth]{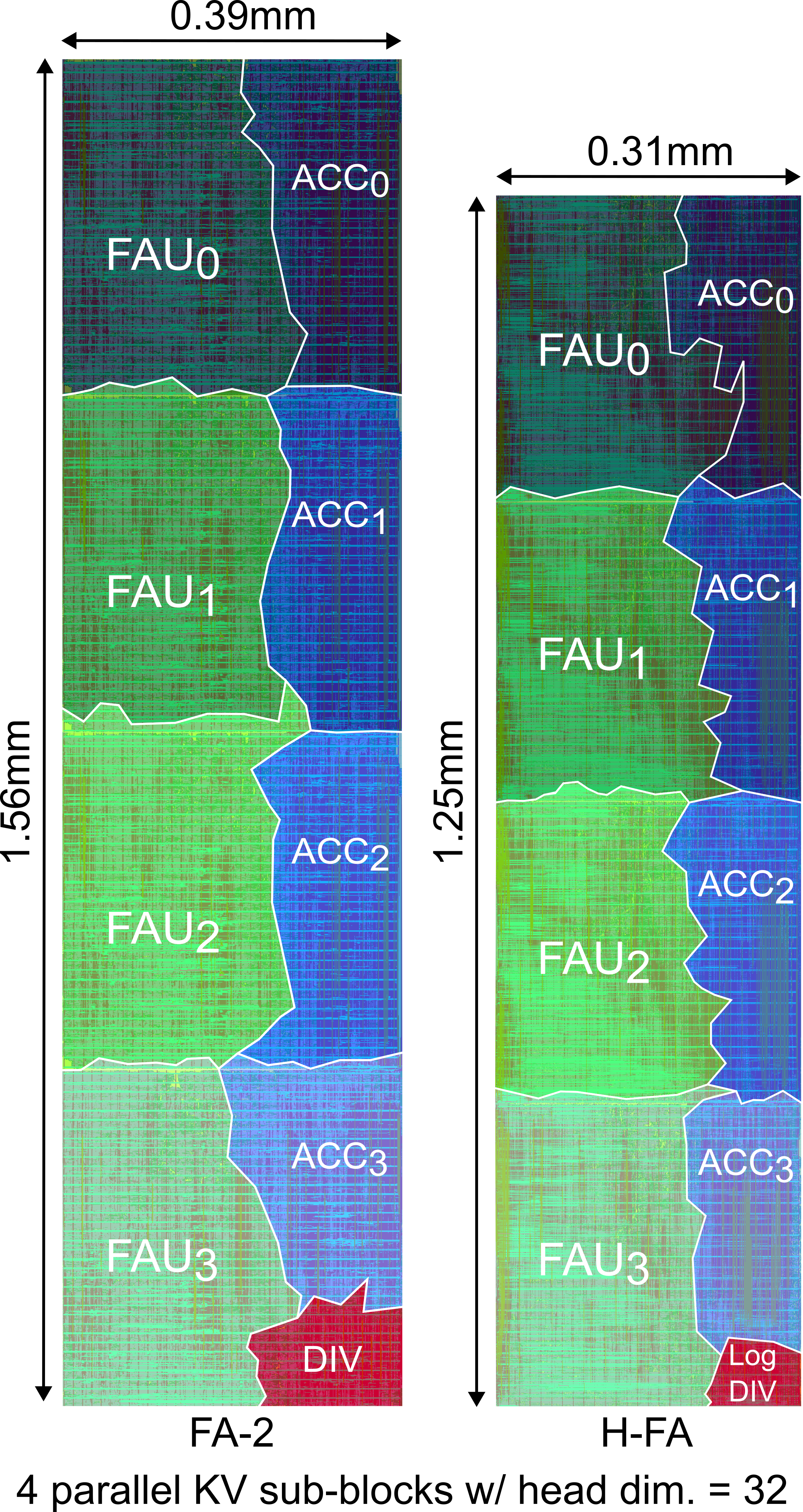}
    \caption{The physical layouts of two attention accelerators following FA-2 (left) and the proposed H-FA (right) architecture. Attention is executed in parallel for one query vector on four KV sub-blocks using four block-FAUs. The partial attention results are merged with four ACC units. The final attention vector for one query is computed at the final division blocks, i.e., DIV and LogDiv for FA-2 and H-FA, respectively.}
    \label{f:phys}
\end{figure}

\begin{figure}[t]
    \centering
    \includegraphics[width=0.8\columnwidth]{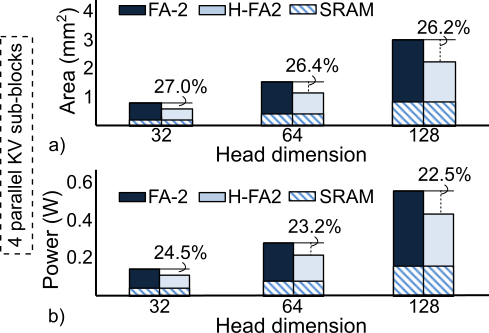}
    \caption{Hardware results for a) area and b) power at 28nm and 500MHz operating frequency for H-FA and FA-2 designs with different head dimension sizes and 4 parallel KV blocks. Both designs utilize the BFloat16 floating point format. Area and power are reported for both datapath and SRAM KV buffers.}
    \label{f:area_power_headsize}
\end{figure}

To better showcase the incurred hardware resource savings, Fig.~\ref{f:phys} highlights the physical layout of the proposed H-FA, relative to the FA-2 design, with 4 parallel KV sub-blocks and a head dimension size of 32.
From the physical layouts, it is evident that H-FA reduces datapath area in both dimensions. The LogDIV block of the H-FA design contains both the subtraction in log domain and the additional logic required for the conversion of the result back to the floating point domain. In total the area reduction is 36.1\% for this case. 
By including the area incurred by the KV buffers area savings are reduced to 27\% as shown in Fig.~\ref{f:area_power_headsize}.

The area and power savings for increasing sizes of head dimension assuming also four KV sub-blocks are summarized in Fig.~\ref{f:area_power_headsize}.
According to Fig.~\ref{f:area_power_headsize}(a) 'H-FA' requires less area than 'FA-2'. The area savings are consistently above 26\% in all examined cases, directly resulting from H-FA's hybrid attention computation, which executes the FlashAttention-2 kernel using both floating-point and logarithmic domains. 
Area and power of SRAM KV buffers are the same for both designs.

\begin{figure}[t]
    \centering
    \includegraphics[width=0.8\columnwidth]{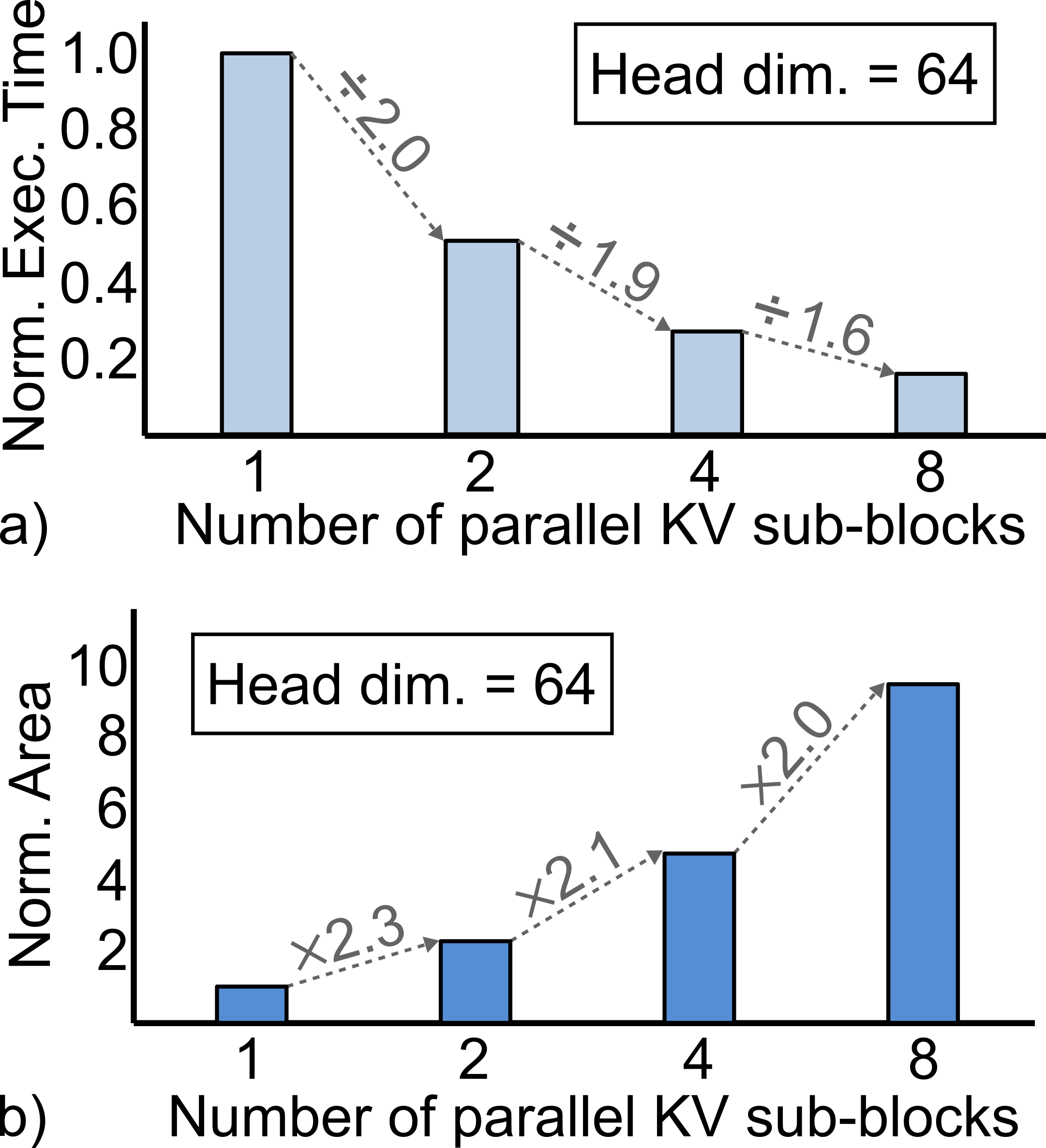}
    \caption{The normalized a) execution time for sequence of $N=1024$ tokens and b) hardware area at 28 nm and 500MHz operating frequency for the a H-FA design with 4 parallel KV sub-blocks and a head dimension size of 64. Both the datapath area and SRAM area were taken into account.}
    \label{f:norm_exec_area}
\end{figure}

\begin{table*}[!h]
\centering
    \caption{Comparison of proposed accelerator and SoTA designs.}
    \label{t:sota}
    \begin{adjustbox}{width=2\columnwidth}
    \centering
    \begin{tabular}{|l||c|c|c|c|c|c|c|c|c|}
    \cline{2-10}
        \multicolumn{1}{c|}{} & \multirow{2}{*}{Platform} & Process & Area & Freq. & Power & \multirow{2}{*}{Precision} & Throughput & Energy Eff. & Area Eff. \\
        \multicolumn{1}{c|}{} & ~ & (nm) & (mm$^2$) & (MHz) & (W) & ~ & (TOPs or TFLOPs) & (TOPs/W) & (TOPs/mm$^2$) \\ \hline
        \multirow{2}{*}{Keller et al.~\cite{keller}} & \multirow{2}{*}{ASIC} & \multirow{2}{*}{5} & \multirow{2}{*}{0.153} & \multirow{2}{*}{152} & \multirow{2}{*}{-} & \multirow{2}{*}{INT4/INT8} & \multirow{2}{*}{3.6/1.8} & \multirow{2}{*}{91.1/39.1} & \multirow{2}{*}{23.53/11.67} \\
        & &  &  & & &  &  & & \\ \hline
        \multirow{2}{*}{MECLA~\cite{mecla}} & \multirow{2}{*}{ASIC} & \multirow{2}{*}{28} & \multirow{2}{*}{22.02} & \multirow{2}{*}{1000} & \multirow{2}{*}{2.87} & \multirow{2}{*}{INT8} & \multirow{2}{*}{14} & \multirow{2}{*}{7.08} & \multirow{2}{*}{0.64}\\
         &  &  &  &  &  &  & &  & \\ \hline
        \multirow{2}{*}{FACT~\cite{fact}} & \multirow{2}{*}{ASIC} & \multirow{2}{*}{28} & \multirow{2}{*}{6.03} & \multirow{2}{*}{500} & \multirow{2}{*}{0.337} & \multirow{2}{*}{INT8} & \multirow{2}{*}{1.02} & \multirow{2}{*}{4.39} & \multirow{2}{*}{0.17}\\
        & & & & & & & & &\\ \hline
        \multirow{2}{*}{Kim et al.~\cite{kim}} & \multirow{2}{*}{ASIC} & \multirow{2}{*}{28} & \multirow{2}{*}{20.25} & \multirow{2}{*}{50} & \multirow{2}{*}{-} & \multirow{2}{*}{INT8} & \multirow{2}{*}{3.41} & \multirow{2}{*}{22.9} & \multirow{2}{*}{0.17} \\
        & & & & & & & & &\\ \hline
        \multirow{2}{*}{Moon et al.~\cite{moon}} & \multirow{2}{*}{ASIC} & \multirow{2}{*}{28} & \multirow{2}{*}{7.29} & \multirow{2}{*}{20} & \multirow{2}{*}{0.002-0.237} & \multirow{2}{*}{AQ 1-8B} & \multirow{2}{*}{0.52} & \multirow{2}{*}{8.94} & \multirow{2}{*}{0.07}\\
        & & & & & & & & &\\ \hline
        \multirow{2}{*}{Chen et al.~\cite{chen}} & \multirow{2}{*}{ASIC} & \multirow{2}{*}{28} & \multirow{2}{*}{0.636} & \multirow{2}{*}{500} & \multirow{2}{*}{0.108} & \multirow{2}{*}{MXINT4/INT8} & \multirow{2}{*}{0.256} & \multirow{2}{*}{2.37} & \multirow{2}{*}{0.40}\\
        & & & & & & & & &\\ \hline
        \multirow{2}{*}{COSA plus~\cite{cosa}} & \multirow{2}{*}{FPGA} & \multirow{2}{*}{16} & \multirow{2}{*}{-} & \multirow{2}{*}{200} & \multirow{2}{*}{30.3} & \multirow{2}{*}{INT8} & \multirow{2}{*}{1.44} & \multirow{2}{*}{0.05} & \multirow{2}{*}{-}\\
         &  &  & & &  &  &  & & \\ \hline 
        \multirow{2}{*}{TSAcc~\cite{tsacc}} & \multirow{2}{*}{ASIC} & \multirow{2}{*}{28} & \multirow{2}{*}{8.6} & \multirow{2}{*}{500} & \multirow{2}{*}{3.1} & \multirow{2}{*}{FP32} & \multirow{2}{*}{2.05} & \multirow{2}{*}{0.66} & \multirow{2}{*}{0.24} \\
         &  &  &  &  &  &  & &  &  \\ \hline \hline
        HFA-1-4 & \multirow{2}{*}{ASIC} & \multirow{2}{*}{28} & \multirow{2}{*}{1.14} & \multirow{2}{*}{500} & \multirow{2}{*}{0.22} & Hybrid & 0.256(BF16)\& & \multirow{2}{*}{5.41} & \multirow{2}{*}{1.02} \\
        (4 KV blocks) &  &  &  & & & BF16\&FIX16 & 0.910(FIX16) & & \\ \hline
        HFA-4-4 & \multirow{2}{*}{ASIC} & \multirow{2}{*}{28} & \multirow{2}{*}{3.34} & \multirow{2}{*}{500} & \multirow{2}{*}{0.62} & Hybrid & 1.64(BF16)\& & \multirow{2}{*}{7.48} & \multirow{2}{*}{1.40} \\ 
        (4 parallel $\vec{q}$, 4 blocks)&  & &  & & & BF16\&FIX16 & 5.84(FIX16) &  &  \\ \hline
    \end{tabular}
    \end{adjustbox}
\end{table*}

Power consumption follows the same trend as shown in Fig.~\ref{f:area_power_headsize}(b). The proposed design requires on average 23.4\% less power. 
Since both designs implement the same FlashAttention algorithm with identical computation order and data flow, SRAM power is 
equivalent in both cases. 

Also, we analyze the trade-off between hardware complexity and runtime improvement as the number of parallel KV sub-blocks increases. Dividing the KV matrices into more sub-blocks increases parallelism, which is expected to reduce overall runtime. However, this parallelism comes at the cost of increased hardware area and power consumption, scaling proportionally with the number of sub-blocks. 

Fig.~\ref{f:norm_exec_area}(a) shows the normalized execution time for a single attention head of 64 elements for $N = 1024$ tokens and b) the corresponding normalized area of the accelerator for the proposed H-FA architecture.

By increasing the number of parallel blocks we are able to achieve significant reduction in kernel execution time, by a factor of $6$ for $8$ parallel KV sub-blocks. Execution time considers only the latency needed to fetch the first query/ies, after that, queries are readily available, through pipelined memory accesses, when the accelerator is ready to start accumulating a new result. As we add more blocks the reaped performance benefits begin to plateau and will eventually come at a significant area overhead cost, since as shown in Fig.~\ref{f:norm_exec_area}(b) area overhead keeps incrementing at approximately the same rate. For the case of $8$ parallel sub-blocks area is increased roughly $10\times$.

\subsection{Comparison with other state-of-the-art designs}
For completeness, we compare the proposed design with other attention accelerators that do not necessarily adopt the fused computation paradigm introduced by the FlashAttention algorithm. For example, state-of-the-art designs used in our comparison rely on generic systolic architectures or multiple vector units for matrix multiplication, while employing custom-designed units to compute non-linear activation functions such as softmax~\cite{chen, keller, kim, moon, mecla, fact,cosa,tsacc}. Among them, only COSA plus~\cite{cosa} and TSAcc~\cite{tsacc} target specifically attention acceleration.

Table~\ref{t:sota} reports comparisons between two configurations of our proposed H-FA accelerator and the selected state-of-the-art designs. The first configuration (named H-FA-1-4) matches the one used in the hardware evaluation of the previous section, where the proposed attention accelerator computes the attention for one query vector using four parallel KV blocks. The second version of the proposed design, named H-FA-4-4 computes the attention of four query vectors in parallel. In the latter case, the datapath of the proposed design is replicated four times, whereas the KV block memory remains shared.

Some designs that accelerate the entire Transformer demonstrate better area and/or power efficiency than the proposed approach. They achieve this by incorporating a large number of processing elements to increase throughput, while keeping power consumption relatively low through reduced hardware complexity, by applying quantization and using narrower datatypes such as INT8.

The proposed designs, which apply logarithmic computation to carefully selected parts of the FlashAttention algorithm, deliver comparable energy and area efficiencies with only a small overhead in area and power. Compared to attention-specific accelerators such as COSA plus~\cite{cosa} and TSAcc~\cite{tsacc}, which also uses similar floating-point datatypes, H-FA-1-4 and H-FA-4-4 achieve significantly higher area and energy efficiencies. COSA plus~\cite{cosa} employs large systolic arrays and can offer significant throughput. However, its energy efficiency is reduced due to its FPGA implementation.

\section{Conclusions}
FlashAttention not only delivers high performance on GPUs but also enables efficient hardware acceleration of attention mechanisms. 
In this work, 
by selectively introducing logarithmic computations in hardware-intensive parts of FlashAttention, we simplify the underlying arithmetic without compromising output accuracy. The proposed architecture implements FlashAttention-2 using a hybrid approach: floating-point arithmetic for attention score computation and fixed-point logarithmic arithmetic for the softmax normalization and value-weighted summation. This design eliminates costly exponentiation, multiplications and division operations, replacing them with lightweight additions and subtractions in the log domain. 
These features lead to substantial reductions in area and power consumption compared to state-of-the-art FlashAttention implementations, while preserving both model performance and the efficient tiling template of FlashAttention. 

\bibliographystyle{IEEEtran}
\bibliography{refs}
\begin{IEEEbiography}[{\includegraphics[width=1in,height=1.25in,clip,keepaspectratio]{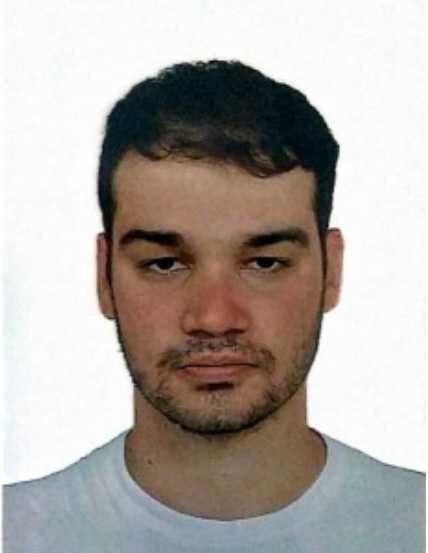}}]{Kosmas Alexandridis} received the Diploma degree in electrical and computer engineering from Democritus University of Thrace, Xanthi, Greece, in 2023, where he is currently pursuing the Ph.D. degree. His research interests include the use of machine learning techniques for verification and design automation of integrated circuits.
\end{IEEEbiography}
\begin{IEEEbiography}[{\includegraphics[width=1in,height=1.25in,clip,keepaspectratio]{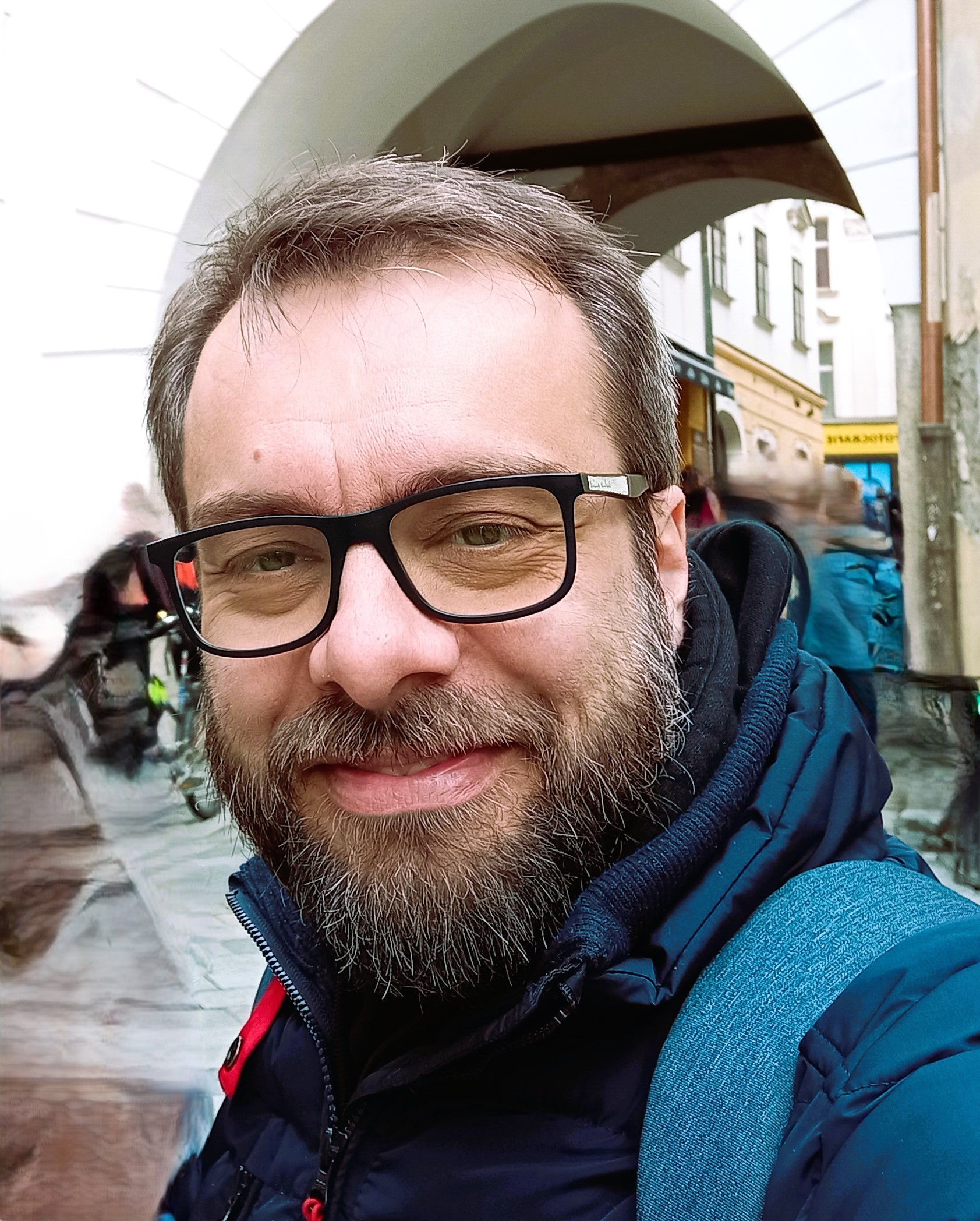}}]{Giorgos Dimitrakopoulos} received the B.S., M.Sc., and Ph.D. degrees in Computer Engineering from the University of Patras, Patras, Greece, in 2001, 2003, and 2007,
respectively. 

He is currently a Professor of Digital Integrated Circuits with the Department of Electrical and Computer Engineering, Democritus University of Thrace, Xanthi, Greece. He is interested in all aspects of digital integrated circuits design and verification. Recently, his research focuses on energy-efficient data-parallel accelerators for machine learning workloads and privacy-preserving technologies, low-cost functional safety, and the use of high-level synthesis for agile chip design.

He received two Best Paper Awards at the Design Automation and Test in Europe (DATE) Conference in 2015 and 2019, respectively. Also, he received the HIPEAC Technology Transfer Award in 2015.
\end{IEEEbiography}
\end{document}